\documentclass[showpacs,nofootinbib]{revtex4}
\usepackage{amsmath}
\usepackage{amsfonts}
\usepackage{amssymb}
\usepackage{dsfont}
\usepackage{mathrsfs}

\newcommand{\source}{{\mathrm{source}}}
\newcommand{\snc}{S_{\nc}}
\newcommand{\nmuno}{(\nc-1)}
\newcommand{\nmdue}{(\nc-2)}
\newcommand{\isovn}{\hat{F}_{\nc}}
\newcommand{\hpsi}{\hat{\Psi}}
\newcommand{\hchi}{\hat{\chi}}
\newcommand{\hphi}{\hat{\phi}}
\newcommand{\bpsi}{\bar{\psi}}
\newcommand{\trans}{{^{\mathrm{T}}}}
\newcommand{\hxi}{\hat{\xi}}
\newcommand{\ha}{\hat{a}}
\newcommand{\had}{\hat{a}^{\dagger}}
\newcommand{\hb}{\hat{b}}
\newcommand{\hbd}{\hat{b}^{\dagger}}
\newcommand{\mm}{\mathcal{M}}
\newcommand{\kin}{-\frac{\hbar^2 \nabla^2}{2m}}
\newcommand{\kinp}{\frac{\hbar^2 k^2}{2m}}

\newcommand{\real}{\mathrm{Re}}
\newcommand{\imag}{\mathrm{Im}}
\newcommand{\en}{i\hbar\partial_{t}}

\newcommand{\nc}{{N_{c}}}
\newcommand{\blom}[4]{\left( 
\begin{array}{ccc} 
#1 & \vdots & #2 \\
\cdots& \cdot & \cdots \\
#3 & \vdots & #4
\end{array}
\right)}
\newcommand{\uno}{\mathbf{1}}

\newcommand{\kkk}{\mathrm{k}}
\newcommand{\xxx}{\mathrm{x}}
\newcommand{\hhh}{\mathrm{h}}
\newcommand{\yyy}{\mathrm{y}}
\newcommand{\zzz}{\mathrm{z}}
\newcommand{\mmm}{\mathfrak{m}}
\newcommand{\nnn}{\mathfrak{n}}
\newcommand{\qqq}{\mathfrak{q}}
\newcommand{\ie}{{\it i.e.}~}

\newcommand{\phin}{{\Phi_{\mathrm{Newtonian}}}}
\newcommand{\vev}{{\it vev}}

\begin{document}
\begin{flushright}
Preprint AEI-2010-167
\end{flushright}

\title{Emergent gravitational dynamics from multi-BEC hydrodynamics?}
\author{Lorenzo Sindoni\footnote{email: \texttt{sindoni@aei.mpg.de}}}
\affiliation{Albert Einstein Institute \\
Am M\"uhlenberg 2, 14476\\
Golm, Germany}
\begin{abstract}
In this paper, we examine the possibility to implement some form of emergent Newtonian gravity in a generic multi-component Bose--Einstein condensate. Parallely to what happens for the emergence of low energy Lorentz invariance, strong requirements have to be imposed on the underlying condensed matter model. We will show, within a simplified model, that the presence of a global symmetry alleviates the problems associated to Lorentz violation, allows the presence of a long range potential, to which the analogue matter fields (the quasi-particles) are coupled following
a weaker form of equivalence principle.
\end{abstract}
\pacs{04.90.+e, 03.75.Kk, 04.60.Bc}
\keywords{}
\maketitle
\setcounter{footnote}{0}
\section{Introduction}

The search for quantum gravity has lead to a variety of different approaches exploring different directions \cite{libro}, with different choices of foundational axioms and with different levels of success. Despite some
promising candidates (like superstring/M-theory \cite{Polchinski} and spinfoam/LQG \cite{Rovelli}) we do not have any completely satisfactory quantum theory for the gravitational field. 

With respect to this quest, the celebrated results coming from the physics 
of black holes in general relativity \cite{Bardeen:1973gs}, and, more recently, derivations of the Einstein equations as equation of state \cite{jacobson} within
a fully thermodynamical treatment (see also \cite{hu,Verlinde,Padmanabhan}) suggest a particular perspective.
In view of these results, the metric would be just a collective
degree of freedom, pretty much like a thermodynamic variable: from this point of view, then, the direct quantization  is not at all the right thing to do. 
While this idea has a certain appeal on its own, it is important to understand what are its limits, and ultimately, what can be said about the properties of the underlying degrees of freedom by the inspection of the properties of the
emergent collective theory.

Analogue models for gravity \cite{analogreview} have provided some useful insights, giving us the opportunity to test with numerical simulations \cite{Carusotto} and even experiments \cite{SilkeHawk,BelgiornoHawk} some key features of physics in strong gravitational fields that would be not so easily accessible otherwise. Additionally, and perhaps most importantly, it has been possible to push a bit further the analysis to include the effects
of some particular trans-Planckian completions on certain predictions for low energy physics \cite{SilkeCosmoPart,CarlosTPlk} and even more exotic phenomena \cite{SilkeSignChg} that have no direct analogue in general relativity.

The power of the analogue models, however, cannot be of great use when we try to address fully dynamical phenomena:
after all, fluid equations of the various sorts cannot be put into one to one correspondence with Einstein equations. 
Nonetheless, while we cannot hope to reproduce dynamical gravitational phenomena with analogue models, we can still
hope to understand some of the symmetry principles that are at the foundations of general relativity. 

The purpose of this paper is to dig in this direction, by considering a particular condensed matter system, namely a Bose--Einstein Condensate (BEC). We will consider the case of a system with several species of bosons, and we will try to address the issue of the emergence of a sensible theory of phonons (which in the analogy will play the role of the matter fields), moving over the condensates (providing the notion of acoustic geometry).
 
Obviously, the most crucial obstacle to the realization of a reasonable theory (within an analogue model) is the absence of symmetries like diffeomorphism invariance and local Lorentz invariance from the very onset. 
As it has been observed \cite{normalmodes}, the most general low energy geometrical structure describing the propagation of perturbations can be cast in the form of a Finsler structure, and not of a Lorentzian one.
Of course, this has deep implications on the possibility of simulating accurately all the phenomena associated to gravity, given the role of local Lorentz invariance in general relativity. Nonetheless, one can hope to get some insight on some structural issues which might be independent from the explicit form (and content) of the equations of motion.

In a previous work \cite{BEC} it has already been discussed the pseudo-gravitational theory that emerges in the case of a single component BEC. Despite being very primitive, it is an example that explicitly shows how, from a microscopic theory (a many-body quantum system), one can obtain some form of (acoustic) metric and a form of pseudo-gravitational theory in the hydrodynamic limit. 
Clearly, despite the obvious limitations of the system, that model is on the right track to implement quantitatively
the otherwise vague idea that general relativity is a hydrodynamic theory.

Despite not being immediately relevant for the discussion of a nonrelativistic theory, to have a form of low energy Lorentz invariance in this kind of models will help in bringing it closer to physically interesting situations.
Therefore we will reconsider the emergence of Lorentz invariance at low energy in BEC, trying to extend (at least partially) the results of \cite{Silke2BEC}, showing that the imposition of an internal symmetry is not enough to
achieve the desired result of a low energy Lorentz invariant effective theory, but that a tuning is still needed.
However, as we will show explicitly, the presence of the underlying symmetry makes Lorentz violation less severe.

For what dynamics is concerned, we will extend the analysis done in \cite{BEC}, considering some backreaction
effects of the phonons onto the condensate. In the case of the single component BEC, it has been shown that
it was possible to mimick some form of nonrelativistic Newtonian dynamics (with important differences that made the
model totally useless for the purpose of realistic dynamical analogues). 
Here we will reproduce the analysis in the more general case, with the objective of finding long range potentials, and, most  importantly, a form of equivalence principle\footnote{We will look, in particular, for universality of free fall.} governing the coupling of matter fields to the candidate for the ``gravitational field'' that we will identify.

The plan of the paper is as follows. In section \ref{sec:2bec} we present a $\mathbb{Z}_2$-symmetric model of a two-components BEC, where we show the emergence of a long range potential. We start the analysis of the general multicomponent case in section \ref{sec:general}, where we discuss the quasi-particle spectrum in homogeneous condensates. Section \ref{sec:beyondGP} deals with the analysis of the dynamics of small inhomogeneities in the condensates as it is induced from the underlying many body Hamiltonian, beyond the Gross--Pitaevski approximation. In the same section, the candidate for the analogue of the Newtonian potential is identified an its relationships with the inhomogeneities in the condensate elucidated. Section \ref{sec:source} is devoted to the discussion of the structure of the source terms for the potentials, while in section \ref{sec:emergent} we discuss the difficulties in obtaining a meaningful local form for the analogue of the Poisson equation for the gravitational field. Section \ref{sec:sn} deals with the generalization of the $\mathbb{Z}_2$ model to a large number of components, partially solving the various difficulties pinpointed in the discussion of the most general case. A summary of the results and further discussions are presented in the concluding section.

\section{A gentle warm up: a 2-components system}\label{sec:2bec}
To clarify the ideas, it is useful to first discuss the simplest instance of a two components systems (similar to the one in \cite{VisserMassiveBEC}), in which we have further simplified the dynamics by imposing a $\mathbb{Z}_{2}$ symmetry. While this symmetry reduces the number of the independent free parameters, it still allows for nontrivial possibilities for the emergent structures. It will also help us to elucidate the pattern that we are following in the general case of many components.

As usual, the system is treated in the second quantization formalism \cite{FetterWalecka}. 
The quantum field equation reads:
\begin{equation}
\en \hpsi_A(\xxx) = \left[ \kin - \mu + \kappa |\hpsi_A|^2 + 
\sigma |\hpsi_B|^2
\right] \hpsi_{A}(\xxx) - \lambda \hpsi_{B} ,
\end{equation} 
while the equation for $B$ is obtained by this by the replacement $A\leftrightarrow B$. The system obtained in this way is $\mathbb{Z}_2$ symmetric.
The masses of the atoms are the same, and as such we have the same chemical potentials, the quartic interaction couplings $\kappa, \sigma$ and the $U(1)$ breaking coupling constant $\lambda$. 

As far as the symmetries are concerned, we see that, of the global $U(1) \times U(1)$ symmetry present if $\lambda = 0$, only a global $U(1)$ survives, corresponding to the invariance over simultaneous phase shifts of the field operators $\hpsi_{A,B} \rightarrow \exp(i\alpha) \hpsi_{A,B}$. According to general arguments (Goldstone's theorem), then, we expect at least one gapless mode, in the spectrum of the excitations.

Condensation, as usual, will be treated within a mean field approximation (MFA).
Of course, this mean field description is just an approximation, and a further analysis is needed in order to test it. In particular, a formalism extending the particle number conserving one \cite{Gardiner,CastinDum}. However, while this
is certainly a point to keep in mind, in the case of almost homogeneous and weakly time dependent configurations, the 
MFA does offer a reliable description, for ordinary BECs. Clearly, this must be checked with care for the general
case we have in mind, but this is a development which is beyond the scope of this paper.
Within the MFA, the second quantized field operators are split into a classical part and a residual fluctuation component as
\begin{equation}
 \hpsi_{X}(\xxx) \approx \psi_{X} \mathbb{I} + \hchi_{A}(\xxx).
\end{equation}
Furthermore, to make comparisons with weak field limit of gravitational theories, we will consider the condensate wavefunctions to be slight deformations of constant configurations 
\begin{equation}
 \psi_{X}(\xxx) = n_{X}^{1/2} \exp(-i\theta_{X}) \left( 1 + \phi_{X}(\xxx) \right),
\end{equation}
where $n_{X}, \theta_{X}$ are constants and $\phi_{X} \ll 1$.
The corresponding Gross--Pitaevski equations read:
\begin{equation}
 \en \psi_A(\xxx) = \left[ \kin - \mu + \kappa |\psi_A|^2 + 
\sigma |\psi_B|^2
\right] \psi_{A}(\xxx) - \lambda \psi_{B}, 
\end{equation}
and the one with $A \leftrightarrow B$.
In the limit of homogeneous condensates, these equations become:
\begin{equation}
 0 = \left[ - \mu + \kappa n_{A} + \sigma n_{B} \right]n_{A}^{1/2} \exp(-i\theta_A) - \lambda_{n_{B}}^{1/2}\exp(-i\theta_B) = 0 
\end{equation}
and the same for the other one, obtained by the replacement $A\leftrightarrow B$. Let us look at the $\mathbb{Z}_2$ symmetric solution:
\begin{equation}
 n_{A} = n_B = n, \qquad \theta_{A} =\theta_{B} = 0,
\end{equation}
where we have used the global $U(1)$ symmetry to fix the phases to be equal to zero.
The equation for $n$, then, gives as a result:
\begin{equation}
 n =\frac{\mu+\lambda}{\kappa+\sigma}.
\end{equation}

The spectrum of quasi-particles around this configuration can be easily derived from the structure of the equation of the fields $\hchi_{X}$, in the limit in which we neglect backreaction effects on the condensate:
\begin{equation}
 \en \hchi_{A}(\xxx) = \left[ \kin -\mu +2\kappa n + \sigma n \right]\hchi_{A} + \kappa n\hchi_{A}^{\dagger} -\lambda \hchi_{B} + \sigma n \hchi_{B} + \sigma n \hchi_{B}^{\dagger}.
\end{equation}
By defining the isovector $ v=\left({\ha_{A}(\kkk)},{\ha_{B}(\kkk)},{\had_{A}(-\kkk)},{\had_{B}(-\kkk)}\right)\trans$
the equation of motion for excitations can be put in the matrix form:
\begin{equation}
 \en v = \left( \kin \mathbb{J} + \mathbb{D} \right) v,
\end{equation}
where
\begin{equation}
 \mathbb{J} = \left( 
\begin{array}{cccc}
1 & 0 & 0 & 0\\
0 & 1 & 0 & 0 \\
0 & 0 & -1 & 0\\
0 & 0 & 0 & -1 
\end{array}
\right),
\end{equation}
and 
\begin{equation}
 \mathbb{D} = \left( 
\begin{array}{cccc}
(2\kappa + \sigma) n - \mu & -\lambda + \sigma n & \kappa n & \sigma n\\
-\lambda + \sigma n & (2\kappa + \sigma) n - \mu & \sigma n &  \kappa n \\
-\kappa n & -\sigma n & -[(2\kappa + \sigma) n - \mu] & \lambda - \sigma n\\
-\sigma n & -\kappa n & \lambda - \sigma n & -[(2\kappa + \sigma) n - \mu] 
\end{array}
\right).
\end{equation}
With a Bogoliubov transformation from the $\ha_{X}$ to new operators $\hb_{X}$
(such that the new operators satisfy the bosonic commutation relations), the equation of motion is diagonalized to the form
\begin{equation}
 \en \tilde{v} = \left( 
\begin{array}{cccc}
\omega_{1}(\kkk) & 0 & 0 & 0\\
0 & \omega_{2}(\kkk) & 0 & 0 \\
0 & 0 & -\omega_{1}(\kkk) & 0\\
0 & 0 & 0 & -\omega_{2}(\kkk) 
\end{array}
\right)
\tilde{v}.
\end{equation}
The eigenfrequencies can be easily computed with this method. They are:
\begin{equation}
 \omega_{1}(\kkk) = \left[ \left(\kinp\right)^2 + 2 (\mu+\lambda) \kinp \right]^{1/2},
\end{equation}
(from which we see that stability requires $\mu+\lambda \geq0$, condition that is equivalent to the positivity of $n$)
and
\begin{equation}
\omega_{2}(\kkk) = \left[ \left(\kinp\right)^2 + 
2\frac{\kappa(3\lambda+\mu)+(\lambda-\mu)\sigma}{\kappa + \sigma} \kinp + 4\lambda \frac{\kappa(2\lambda+mu)-\mu \sigma}{(\kappa + \sigma)} \right]^{1/2} .
\end{equation}
We see that the second mode has a nonzero gap, and therefore will represent the massive boson we are interested in. The massless mode is the Goldstone boson associated to the breaking of the global $U(1)$ symmetry changing both the phases of the field operators $\hpsi$. Notice also that the gap is proportional to $\lambda$. In fact, this mode is the pseudo-Goldstone boson associated to the soft breaking of the global $U(1)$ symmetry, which is broken precisely by the coupling term (parametrized by $\lambda$) between the two species.

If we want to recover the special relativistic spectrum, at least at low energies, we need the coefficients of the quadratic terms in $k$ in the two expressions above to coincide:
\begin{equation}
(\kappa + \sigma) (\mu + \lambda) =\kappa(3\lambda+\mu)+(\lambda-\mu)\sigma,
\end{equation}
leading to
\begin{equation} \label{mono2}
 \lambda = \frac{\sigma}{\kappa} \mu.
\end{equation}
Notice that $\sigma = 0$ leads to $\lambda = 0$ which is another condition in which we have monometricity, but is also the case in which the two condensates are perfectly decoupled.

In the case where \eqref{mono2} is satisfied, the two spectra are:
\begin{equation}
 \omega_{1}(\kkk) = \left[ \left(\kinp\right)^2 + 2 (1+x)\mu \kinp \right]^{1/2},
\end{equation}
\begin{equation}
 \omega_{2}(\kkk) = \left[ \left(\kinp\right)^2+ 2 (1+x)\mu\kinp + 4 \mu^2 x\right]^{1/2},
\end{equation}
where we have introduced the dimensionless ratio:
\begin{equation}
 x = \frac{\sigma}{\kappa}.
\end{equation}
The healing scale is computed in the usual way \cite{FetterWalecka}. It turns out to be:
\begin{equation}
 L_{heal}^2 = \frac{1}{1+x} \frac{\hbar^2}{4m\mu}.
 \end{equation}
 
Even more important than the spectrum, at least in this case, is the exact form of the Bogoliubov transformation. In principle, the calculation can be carried on with the full $k-$dependence of the various quantities.
However, the detailed discussion of these transformations lays beyond the limits of this paper. The only thing that we need to remember about them is that they are nonlocal field transformations defining the quasi-particle fields (associated to the physically propagating {\it collective} modes) in terms of the fluctuation operators $\hchi_X$.

In order to understand the coupling between the phonons and the ``gravitational field'' we need to write the field equation for $\hchi_X$ when the background condensate is slightly inhomogeneous:
\begin{equation}
 \en \hchi_{A} = \left[ \kin - \mu +2\kappa |\psi_A|^2 + \sigma |\psi_B|^2\right] \hchi_A + 
\kappa \psi_A^2\hchi_{A}^\dagger
+ \lambda\hchi_B + \sigma \psi_{A}\psi_B \hchi_B^{\dagger} + \sigma \psi_{A}\bar{\psi}_{B} \hchi_{B},
\end{equation}
where now
\begin{equation}
 \psi_{X} = n^{1/2}(1+u_X+iv_X).
\end{equation}
We see that the terms coupling the backgrounds $u_X,v_X$ to the quasi-particles are:
\begin{equation} \label{couplings}
 2n (\kappa u_A +\sigma u_B) \hchi_{A} + 2\kappa n (u_A+i v_A) \hchi_A^{\dagger} + \sigma n ((u_A+u_B)+i(v_A+v_B))\hchi_{B}^{\dagger} + \sigma n [(u_A+u_B)+i(v_A-v_B)]\hchi_{B}.
\end{equation}
From this expression one easily understand that the effect of the inhomogeneities, from the point of view of quasi-particles, is not only a scattering off a potential, but it can also be a transition to a different color. Instead of looking for more details, which will be carefully examined later, we will pass to the analysis of the potentials.

\subsection{The potentials}
As we have seen before, the quasi-particles will be coupled to the inhomogeneities of the condensates in a variety of ways. It is important to distinguish among them, the one that is going to play the role of the ``Newtonian potential''. Therefore we need to understand what happens to the Gross--Pitaevksi equation when we insert explicitly nonhomogeneities. It is immediate to see that the equation will have the following form:
\begin{equation}
 0 = \left[ \kin - \mu + (2\kappa + \sigma) n  \right] \phi_{A} + (\sigma n - \lambda)\phi_{B} +\kappa n \bar\phi_{A}+
\sigma n \bar\phi_{B}  + \source,
\end{equation}
where $\source$ denotes various expectation values generalizing the anomalous mass and the anomalous densities of the single BEC.
Taking the real and imaginary parts, we can put the equation in the form:
\begin{equation}
\left( \kin \mathbb{I} + \mu \left(
\begin{array}{cccc}
 (2+x) & x & 0 & 0 \\
x & (2+x) & 0 & 0 \\
0 & 0 & x & -x \\ 
0 & 0 & -x & x \\
\end{array}
 \right) \right) \left(\begin{array}{c}
 u_A \\ u_B \\ v_A \\ v_B
\end{array}
\right) = \source.
\end{equation}

This matrix is block diagonal. While the modes arising from the $u$-fields are short range, among the ones arising from the $v-$fields there is one that is long range, as a direct calculation can show.
In particular, the mode:
\begin{equation}
V_{long} = v_{A} + v_{B} 
\end{equation}
will obey an equation of motion of the form:
\begin{equation} \label{equlong}
-\frac{\hbar^{2} \nabla^{2}}{2m} V_{long} = \source,
\end{equation}
which has to be compared with the Poisson equation for the gravitational field:
\begin{equation}
\nabla^{2} \Phi_{N} = 4 \pi G \rho_{matter}.
\end{equation}
The correct mapping between $V_{long}$ and $\Phi_{N}$ is just a matter of keeping track of all the dimensionfull coefficients appearing in the equations \eqref{equlong} and \eqref{couplings} (see \cite{BEC} for the single component case).
In conclusion, in this simple model, there is room for emerging Lorentz invariance at low energy for all the particles, while there is simultaneously the possibility to coupling the massive mode to a long range nonrelativistic potential.

We have used the most symmetric Lagrangian that allows for at least one massive mode. One could have obtained monometricity without tuning of parameter by enhancing the symmetry from $\mathbb{Z}_2$ to $SU(2)$ by setting $\sigma = \kappa$ and $\lambda=0$. The price to pay is that in this model, as one can see, there are no gapped modes, and therefore this extended symmetry excludes the possibility of discussing the nonrelativistic limit of a massive relativistic particle.

Notice also that the presence of a symmetry in the fundamental action is not enough to enforce monometricity. Within all the symmetric Lagrangians compatible with the requirements, one has still has to tune the parameters of the model \cite{Silke2BEC}. This fact will carry over in the general multi-component case
that we are going to discuss now.

\section{The general multi-component model}\label{sec:general}
Having examined in a certain detail the case of a two-component system, we can now give the description of the general multi-component case.
Every component is associated to a second quantized field operator,
\begin{equation}
\hpsi_{A}(\xxx) = \frac{1}{\sqrt{V}} \sum_{k} \hat{a}_{A}(\kkk)e^{i\kkk\cdot \xxx},
\end{equation}
where $A=1,...,N_{c}$ is labeling the components. For convenience, we will often refer to the index $A$ as a color index, even though it is not necessarily associated to a representation of some internal symmetry group. For instance, it could be associated to a spin component as in the so-called spinor condensates \cite{spinorBEC} or to some more general hyperfine level like in two components BECs.

The Hamiltonian of the system is given as:
\begin{equation}
\hat{H} = \int d^{3}x \left( -\sum_{A} \hpsi_{A}^{\dagger}(x) \frac{\hbar^{2}\nabla^{2}}{2m_{A}} \hpsi_{A}(x) + V(\hpsi) \right),
\end{equation}
where $m_{A}$ denotes the mass of the atoms of the $A$-th color and $V$ is a potential term controlling the interactions among the particles. The general structure of this term is not specified. In the rest of the paper we will try to establish a number of conditions which must be imposed onto the potential in order to match the properties of a Newtonian theory of gravity.

\subsection{Condensation}
As usual the condensation is described with the mean field approximation (MFA):
\begin{equation}
\hpsi_{A}(\xxx)\approx \psi_{A}(\xxx)\mathbb{I} + \epsilon \hchi_{A}(\xxx),
\end{equation}
where $\psi_{A}$ represents the condensate wavefunction associated to the color $A$, $\hchi_{A}$ represents the corresponding fraction over the condensate and $\epsilon\ll1$ has been introduced in order to explicitly keep into account the fact that the noncondensed fraction is small. It might be thought as a $1/N$ term, where $N$ is the total number of particles, if one imagines a sort of asymptotic expansion in the number of particles.

In the limit of $\epsilon\rightarrow 0$, the condensate and the noncondensate parts can be thought as weakly coupled (in particular the ``backreaction'' of the noncondensate fraction on the condensate can be neglected), and the analysis of some physical features of the system are easier to be described. In particular, the equation of motion for the field operator becomes an equation of motion for the classical fields $\psi_{A}$:
\begin{equation}
i\hbar \frac{\partial \psi_{A}}{\partial t} = -\frac{\hbar^{2} \nabla^{2}}{2m_{A}} \psi_{A} + \frac{\partial V}{\partial \psi^{*}_{A}}(\psi). 
\end{equation}
These equations are the generalization of the Gross--Pitaevski equation to the case of the multi-component condensate. Among the solutions, we are interested into the constant ones, associated to a flat background acoustic geometry. We will denote them by $\bpsi_{A}$. They solve the equations:
\begin{equation}
\frac{\partial{V}}{\partial \psi^{*}_{A}} (\bpsi) =0,
\end{equation}
and their complex conjugates.

The number of these constant solutions depends on the number of the minima of the potential $V$. Symmetries of the potential might also affect the structure of the set of these ``vacuum states''. The detailed discussion of these vacua is immaterial for our purposes (at least at this stage), and it is sufficient to focus on a single configuration $\bpsi_{A}$ without digging the full structure of the possible configurations.  

Before passing to the discussion of the quasi-particles, it is useful to give some relations between the various derivatives of the potential with respect to the field arguments. In order for the quantum system to be consistent, besides the positivity of energy which is ensured by the shape of the potential $V$, the Hamiltonian must be an Hermitian operator. Consequently, the potential must be an Hermitian operator. When translated in terms of a complex function of complex variables, the hermiticity constraint can be written as:
\begin{equation}
V(\psi,\psi^{*}) = V(\psi^{*},\psi)^{*}.
\end{equation}
Therefore, when taking derivatives with respect to the arguments, we get:
\begin{equation}
\frac{\partial V}{\partial \psi_{A}} = \left(\frac{\partial V}{ \partial \psi^{*}_{A}}\right)^{*},
\end{equation}
\begin{equation}
\frac{\partial^{2} V}{\partial \psi_{A}\partial \psi_{B}} = \left(\frac{\partial^{2} V}{ \partial \psi^{*}_{A} \partial \psi^{*}_{B}}\right)^{*},
\qquad
\frac{\partial^{2} V}{\partial \psi_{A}\partial \psi^{*}_{B}} = \left(\frac{\partial^{2} V}{ \partial \psi^{*}_{A} \partial \psi_{B}}\right)^{*}.
\end{equation}
These identities will play a role in slightly simplifying the following analysis.

\subsection{Quasiparticles} \label{subsec:genquasip}
In the expansion in power of $\epsilon$, while the lowest order equations are giving the GP equations for the condensate wave-functions, the next order is giving the equation for the propagation of the fields $\hchi_{A}$.
These equations are:
\begin{equation}
i\hbar \frac{\partial \hchi_{A}}{\partial t} = - \frac{\hbar^{2}\nabla^{2}}{2m_{A}} \hchi_{A} + X_{AB}\hchi_{B}+
Y_{AB} \hchi^{\dagger}_{B} ,
\end{equation}
where
\begin{equation}
X_{AB}=\left. \frac{\partial^{2} V}{\partial \psi_{A}\partial \psi_{B}} \right|_{\psi=\bpsi},\qquad
Y_{AB}=\left.\frac{\partial^{2} V}{\partial \psi_{A}\partial \psi^{*}_{B}}\right|_{\psi=\bpsi}
\end{equation}
are constant matrices. They are responsible for mixing between different colors as well as for mixing between particle and hole states. The propagating modes, the quasi-particles, must be obtained by a suitable Bogoliubov transformation of the original modes into eigenstates of the Hamiltonian. Without entering into the details of the calculation, the general relation between the quasi-particle field operators $\hphi_{A}$ and the particle field operators can be written in a very general way as:
\begin{equation}
\hphi_{A}(x) = \int d^{3}y \left( f_{AB}(x,y) \hchi_{B}(y) + g_{AB}(x,y) \hchi^{\dagger}_{B}(y)  \right),
\end{equation}
where the kernels $f_{AB},g_{AB}$ are calculable directly from the Bogoliubov coefficients usually
computed in momentum space\footnote{For a detailed discussion of the implications, see \cite{BEC,wroclaw}.}. The diagonalization procedure leads also to the spectrum of the quasi-particles,
\begin{equation}
i\hbar \frac{\partial \hphi_{A}}{\partial t} = E_{A}(-\nabla^{2}) \hphi_{A} + E^{(0)}_{A}\hphi_{A},
\end{equation}
where the term $E^{(0)}_{A}$ has been introduced in such a way to explicitly show the energy gap of the mode. In terms of the wave number $k$, the energy of the mode is
\begin{equation}
E_{A}(k) = E_{A}(k) + E^{(0)}_{A} , \qquad E_{A}(0) = E^{(0)}_{A}.
\end{equation}
The presence or the absence of a gap is related to particular symmetries of the Hamiltonian and of the background solutions $\bpsi_{A}$.
For instance, if some global symmetry is spontaneously broken by the choice of $\bpsi$, then some gapless Goldstone bosons will necessarily arise.

In order to translate the gaps into rest masses, a notion of speed of sound is needed. Therefore, the structure of the dispersion relations of the quasi-particles must be investigated in more details.
At low momenta, the dispersion relation can be expanded in powers of $p=\hbar | \vec{k}|$:
\begin{equation}
E_{A}(p) \approx e_{1,A} p + e_{2,A} p^{2}. 
\end{equation}
If the mode is gapped, in order to have a similarity with non-relativistic particles, the coefficients $c_{1,A}$ must vanish:
\begin{equation}\label{eq:lineardisp}
e_{1,A}=0.
\end{equation}
Obviously, these are conditions which are imposed upon $\bpsi$ (and hence indirectly on the potential $V$). There are as many conditions as the number of gapped modes. 
Furthermore, the second term in the expansion must be read as a nonrelativistic kinetic term:
\begin{equation}
e_{2,A} p^{2} = \frac{p^{2}}{2 \mm_{A}},
\end{equation}
where $\mm_{A}$, which is defined by this equation, is the effective mass for the quasi-particle $A$. With these definitions one can provide a notion of ``speed of sound''\footnote{This definition is only formal. The correct procedure should involve the discussion of the group velocity in the relativistic regime. However, this is not available for the present discussion.} by means of the gap energy:
\begin{equation}
E^{(0)}_{A} = \mm_{A} c_{s,A}^{2} \Rightarrow c_{s,A}^{2} = \frac{E^{(0)}_{A}}{\mm_{A}}.
\end{equation}
Monometricity conditions are then equivalent to the requirements:
\begin{equation}\label{eq:monometricity}
c_{s,A} ^{2} = c^{2}\,, \forall A.
\end{equation}
They represent a certain number of constraints on the potential $V$ (and, of course, on the solutions $\bpsi_{A}$). It is rather complicated to disentangle them and write down a simple condition on the potential $V$. For the purposes of this paper it will be enough to leave them in this implicit form. For a detailed discussion of this problem in the case of the 2BEC system see \cite{Silke2BEC}.

Here we find the first tuning issue. As in the case of 2BECs, the monometricity of the emergent acoustic structure is
far from obvious. In particular, there are distinct low-energy geometric phases (monometric, multimetric and genuinely Finslerian \cite{SilkeLNP}),
each of them corresponding to a given region of the parameter space for the microscopic parameters.

\section{Beyond Gross--Pitaevski}\label{sec:beyondGP}

So far, the discussion involved just the homogeneous solutions, where the backreaction effects due to quasi-particles are completely discarded. To include them, it is needed to change slightly the formalism and to pass to the Bogoliubov--de Gennes one. This is still a MFA method, but now the effect of the noncondensed fraction is systematically introduced and taken into account. As for the Gross--Pitaevski equation, there is a straightforward generalization to be done in order to
include the various colors. 
A consequence of the mean field approximation is that 
\begin{equation}
\langle \hchi_{A} \rangle = 0,
\end{equation}
so the first nonvanishing expectation values are the quadratic ones:
\begin{equation}
\mmm_{AB}(\xxx) = \langle \hchi_{A}(\xxx) \hchi_{B}(\xxx) \rangle, 
\qquad \nnn_{AB}(\xxx) = \langle \hchi_{A}^{\dagger}(\xxx) \hchi_{B}(\xxx) \rangle, 	
\qquad \qqq_{AB}(\xxx) =\langle \hchi_{A}^{\dagger}(\xxx) \hchi_{B}^{\dagger}(\xxx) \rangle,
\end{equation}
which are generalizing the anomalous mass and density of the single-BEC system. A normal ordering with respect to the basis $\hchi$ removes an unphysical zero point energy\footnote{Notice that this normal ordering is not equivalent to normal ordering with respect to the quasi-particle field operators $\hphi$.}, as well as terms of the form $\langle \hchi \hchi^{\dagger} \rangle$.

\subsection{Condensate}
The general study of the properties of nonhomogeneous condensates is a topic which is beyond the scope of
this paper. 
Rather, it will be enough to restrict the analysis to small deviations from homogeneity:
\begin{equation}
\psi_{A} \approx \bpsi_{A} + u_{A}(x) + i v_{A}(x),
\end{equation}
which corresponds, as we have already mentioned, to the weak field regime of the corresponding acoustic geometry. The slight discrepancy with the notation used in section \ref{sec:2bec} is due to convenience.

Using the MFA into the equation for the evolution of the operator $\hpsi$ and taking the expectation value, we obtain an equation for the condensate which explicitly takes into account the (lowest order) contribution of the noncondensed fraction. If one neglects higher order expectation values, which can be thought to be further suppressed by powers of $1/\sqrt{N}$ with respect to the anomalous masses and densities, one finds the following equations:
\begin{equation}\label{eq:perturbationsgeneral}
i\hbar\frac{\partial}{\partial t}(u_{A} + iv_{A}) = \frac{-\hbar^{2} \nabla^{2}}{2m_{A}} (u_{A}+iv_{A}) + X_{AB}(u_{B}+iv_{B}) + Y_{AB} (u_{B}-iv_{B}) +\Sigma_{A},
\end{equation}
where $\Sigma_{A}= \Sigma_{A}^{R} + i \Sigma_{A}^{I}$ are the source terms encoding the effect of fluctuations around the mean field,
\begin{equation}
\Sigma_{A} = \left.\frac{\partial X_{AB}}{\partial \psi_{C}} \right|_{\psi=\bpsi} \mmm_{BC} +
\left.\left(\frac{\partial X_{AB}}{\partial \psi^{*}_{C}} + \frac{\partial Y_{AB}}{\partial \psi_{C}} \right)\right|_{\psi=\bpsi} \nnn_{BC} +
 \left.\frac{\partial Y_{AB}}{\partial \psi^{*}_{C}} \right|_{\psi=\bpsi} \qqq_{BC}.
\end{equation}

Taking the real and imaginary parts of the Eq. \eqref{eq:perturbationsgeneral}, and by considering quasistatic configurations, for which time derivatives give sub-leading contributions, the equations for the fields $u_{A}$ and $v_{A}$ are easily obtained
\begin{equation}
 \frac{\hbar^{2} \nabla^{2}}{2m_{A}} u_{A} - \Omega_{AB} u_{B} - \Xi_{AB} v_{B} = \Sigma_{A}^{R},
\end{equation}
\begin{equation}
 \frac{\hbar^{2} \nabla^{2}}{2m_{A}} v_{A} - \Upsilon_{AB} v_{B} - \Gamma_{AB} u_{B} = \Sigma_{A}^{I},
\end{equation}
where, for convenience, we have introduced the notation
\begin{equation}
\Omega_{AB} = \real(X_{AB}+Y_{AB}), \qquad \Xi_{AB} = \imag(X_{AB}-Y_{AB}),
\end{equation}
\begin{equation}
\Upsilon_{AB} = \real(X_{AB}-Y_{AB}), \qquad \Gamma_{AB} = \imag(X_{AB}+Y_{AB}).
\end{equation}
These matrices control certain mixing terms between the different perturbations of the condensates, which are the candidates for representing our emergent interactions. These mixing terms must be 
reabsorbed by a suitable diagonalization, in such a way that we are left with equations in the form:
\begin{equation}
\nabla^{2} \tilde{u}_{A} + \frac{1}{L_{A}^{2}} \tilde{u}_{A}= \source.
\label{eq:pseudopots}
\end{equation}
However, for this diagonalization to be performed, some conditions must be met. In particular, let us introduce the matrix $M=\mathrm{diag}(m_{1},...,m_{N_{c}})$ and the
the two $2N_{c}\times 2N_{c}$ matrices,
\begin{equation}
\mathds{A}=\left( \begin{array}{cc} M & 0 \\ 0 & M \end{array} \right), \qquad \mathds{B}= \left( \begin{array}{cc} \Omega & \Xi \\ \Gamma & \Upsilon \end{array} \right),
\end{equation}
such that the equations can be written as
\begin{equation}
 \frac{\hbar^2}{2}\nabla^2 ( \mathds{A}.w) + \mathds{B}.w = 0,
\end{equation}
where $w=(u_1,...,u_{N_c},v_1,...,v_{N_c})\trans$.

Then, the condition we are seeking is that $\mathds{A}$ and $\mathds{B}$ must be simultaneously diagonalizable. Since $M$ is already diagonal, this poses strong constraints on the matrices $\Xi,\Gamma,\Upsilon,\Omega$,
unless the degeneracy of the eigenvalues of $M$ is large. If $M = m \mathbb{I}$, then the only remaining condition is that $\mathds{B}$ is symmetric\footnote{We have turned our complex equations into real equations in twice the number of variables. Therefore the matrices we are playing with are real.}.
This provides three matrix conditions on each block:
\begin{equation}
\Omega^{T}=\Omega, \qquad \Upsilon^{T} = \Upsilon, \qquad \Gamma^{T} = \Xi,
\end{equation}
which are ultimately conditions on the second derivatives of the potential $V$, evaluated on the background solution $\bpsi_{A}$. If these are not met, then the LHS of the equations cannot be put into diagonal form and there is an unavoidable mixing between the various potentials. 

However, it is easy to see that the very definition of the matrices $X_{AB}$ and $Y_{AB}$ (and in particular the condition that the Hamiltonian is Hermitian) ensures that these conditions are identically satisfied. Indeed, if $V$ is sufficiently smooth, 
\begin{equation}
X_{AB} = \frac{\partial^{2} V}{\partial \psi_{A}\partial \psi_{B}} = \frac{\partial^{2} V}{\partial \psi_{B}\partial \psi_{A}} = X_{BA},
\end{equation}
which ensures that $X_{AB}$ is a symmetric matrix, while
\begin{equation}
Y_{AB} = \frac{\partial^{2} V}{\partial \psi_{A}\partial \psi^{*}_{B}} = \left(  \frac{\partial^{2} V}{\partial \psi^{*}_{A}\partial \psi_{B}} \right)^{*} = Y_{BA}^{*},
\end{equation}
which implies:
\begin{equation}
\real({Y_{AB}}) = \frac{Y_{AB}+Y_{AB}^{*}}{2} = \frac{Y_{AB}+Y_{BA}}{2} = \real({Y_{BA}}),
\end{equation}
\begin{equation}
\imag(Y_{AB})=\frac{Y_{AB}-Y_{AB}^{*}}{2i} = \frac{Y_{AB}-Y_{BA}}{2i} = - \imag({Y_{BA}}).
\end{equation}
Using these properties it is immediate to prove that $\Omega_{AB}=\Omega_{BA}$ and $\Upsilon_{AB}=\Upsilon_{BA}$, from the symmetry of the matrices 
$X_{AB}$ and $\real{(Y_{AB})}$, while the antisymmetry of the matrix $\imag(Y_{AB})$ ensures that $\Gamma_{BA}=\Xi_{AB}$. Therefore, the mixing matrix between the various potentials is symmetric, and hence diagonalizable. 
Consequently, the diagonalizability condition 
\begin{equation}\label{eq:diagonalizability}
\left[\mathds{A},\mathds{B}\right]=0
\end{equation}
can be seen ultimately as a condition on the degeneracy of the matrix $M$.

If the diagonalization can be performed, then the equations take the form:
\begin{equation}\label{eq:diagonalpoisson}
-\frac{\hbar^{2}}{2m_{A}}\nabla^{2} \tilde{u}_{A} + \lambda_{A,u} \tilde{u}_{A}= p^{B}_{A} \Sigma_{B}^{R} + q^{B}_{A} \Sigma_{B}^{I},
\end{equation}
where $p,q$ are the matrices related to some blocks of the diagonalization matrix, $m_{A}$ are the eigenvalues of the mass matrix, and $\lambda_{A}$ are eigenvalues of the matrix $\mathds{B}$. Of course, analogous equations hold for  fields $\tilde{v}_{A}$.

It is the case to spend few words concerning the necessity of this diagonalization. Of course, from the point of view of the physics, the potentials $u,v$ have
the most immediate relevance, being directly related to fluid dynamical variables of the condensate (via the Madelung representation). Nonetheless, $\tilde{u},\tilde{v}$ represent the modes (of the condensate) that can be excited independently (at least at the lowest order in perturbation theory). This is why they are so important.

The terms $\lambda_{A}$ are related to length scales through:
\begin{equation}
\frac{1}{L_{A}^{2}} = \frac{2\lambda_{A}m_{A}}{\hbar^{2}}.
\end{equation}
In the case of the single BEC, this turned out to be the healing length, which represents the only dynamical scale present in the game. This scale is responsible for the extremely short range potential emerging in that example. However, having at our disposal more components, and in particular several interaction terms between them, it happens that more interesting situations can be realized. While a general rule is impossible to be discussed, it is illuminating to show what could happen in a simplified case. 
Let us assume that for some reason, related to the symmetry structure of the system, as in section \ref{sec:2bec}, one ends up with a set of coupled equations like:
\begin{equation}
\nabla^{2}\left(\begin{array}{c}u_{1}\\u_{2}\end{array} \right) + \left(\begin{array}{cc}a & a\\a&a\end{array} \right)\left(\begin{array}{c}u_{1}\\u_{2}\end{array} \right) = \source,
\end{equation}
then, diagonalizing, one finds:
\begin{equation}
\nabla^{2}\left(\begin{array}{c}\tilde{u}_{1}\\ \tilde{u}_{2}\end{array} \right) + \left(\begin{array}{cc}2a & 0\\0&0\end{array} \right)\left(\begin{array}{c}\tilde{u}_{1}
\\ \tilde{u}_{2}\end{array} \right) = \source,
\end{equation}
where one of the two potentials is short range, while the other is long range, giving rise to $1/r$ potentials, rather than Yukawa ones. Therefore, while the interaction between components complicates the analysis of the physical properties, it opens the possibility of creation of long range potentials which are strictly forbidden in a single component BEC. However, as one can imagine, this feature is not generic at all. Some requirements are needed concerning the (global)
symmetries of the system on one hand, and on the symmetries that are preserved by the particular choice of the vacuum state on the other. In particular, a situation like the one briefly discussed here will be realized in the system considered in section \ref{sec:sn}.

\subsection{Quasiparticles}

To study the properties of quasi-particles, in principle one should go for the Bogoliubov transformation 
in the case of non-homogeneous backgrounds. 
First of all, rewrite the equation for the excitations in non-homogeneous backgrounds as
\begin{equation}
 i\hbar \frac{\partial}{\partial t} \left( \begin{array}{c}
 \vec{\hchi}(\xxx) \\ \vec{\hchi}(\xxx)^{\dagger} 
\end{array}
 \right) = \mathcal{G} \left( \begin{array}{c}
 \vec{\hchi}(\xxx) \\ \vec{\hchi}(\xxx)^{\dagger} 
\end{array}
 \right) + \mathcal{U}(\xxx) \left( \begin{array}{c}
 \vec{\hchi}(\xxx) \\ \vec{\hchi}(\xxx)^{\dagger} 
\end{array}
 \right),
\end{equation}
where $\mathcal{G}$ denotes both the kinetic operator and the linear term for the homogeneous background while the $\mathcal{U}$ matrix encodes the inhomogeneities alone, which are contained just at the linear order, in the approximation scheme we are using.
Doing the Fourier transform with respect to the modes $e^{-ik\xxx}$, we get
\begin{equation}
  i\hbar \frac{\partial}{\partial t} \left( \begin{array}{c}
 \vec{\hat{a}}(k) \\ \vec{\hat{a}}^{\dagger} (-k) 
\end{array}
 \right) = \mathcal{G}(k) \left( \begin{array}{c}
 \vec{\hat{a}}(k) \\ \vec{\hat{a}}^{\dagger} (-k) 
\end{array}
 \right) + \int d^3x \, e^{-ik\xxx} \left[ \mathcal{U}(\xxx) \left( \begin{array}{c}
 \vec{\hchi}(\xxx) \\ \vec{\hchi}(\xxx)^{\dagger} 
\end{array}
 \right)\right],
\end{equation}
The last term can be manipulated by introducing the Fourier transforms:
\begin{equation}
 \mathcal{U}(\xxx) = \int d^3 h \, e^{ih\xxx} \tilde{ \mathcal{U}}(h),
\end{equation}
to get, finally
\begin{equation}
\label{eq:inhombogoliubov}
  i\hbar \frac{\partial}{\partial t} \left( \begin{array}{c}
 \vec{\hat{a}}(k) \\ \vec{\hat{a}}^{\dagger} (-k) 
\end{array}
 \right) = \mathcal{G}(k) \left( \begin{array}{c}
 \vec{\hat{a}}(k) \\ \vec{\hat{a}}^{\dagger} (-k) 
\end{array}
 \right) + \int d^3h \,\tilde{\mathcal{U}}(\hhh) \left( \begin{array}{c}
 \vec{\hat{a}}(\kkk-\hhh) \\ \vec{\hat{a}}^{\dagger}(\hhh-\kkk) 
\end{array}
 \right).
\end{equation}
This means that the Bogoliubov transformations, in this case, are much more complicated and involve different momenta, according to the spectrum of $\mathcal{U}(\xxx)$. This is expected, since the simplicity of the Bogoliubov transformation for the homogeneous background was deeply related to translational invariance, which is broken by the presence of the inhomogeneities.

Given the intrinsic difficulties in disentangling Eq.\eqref{eq:inhombogoliubov}, we will not try to compute the Bogoliubov transformations in this case. Rather, we will try to deduce the Feynman rules for the interaction between phonons and the inhomogeneities, using a scattering theory point of view.

We will use, then, the free states obtained linearizing around the homogeneous background, \ie the phonons described in section \ref{subsec:genquasip}, without including the effect of the corrections to Bogoliubov transformations. We will treat these states as asymptotic states, in the sense of scattering theory, imagining that the inhomogeneities are small {\it localized} ripples in the condensate. This is another limitation of this analysis but, nonetheless, we will be able to grasp the essential features of the physics of the system.

The perturbation in the Hamiltonian for the quasi-particles is encoded in the operator
\begin{equation}
 \hat{H}_{int} = \int d^3 x \left( a^{AB}(x) \hchi_A(\xxx)\hchi_B(\xxx) +b^{AB}(x) 
\hchi^\dagger _A(\xxx)\hchi_B(\xxx)+
\,h.c.\,\right),
\end{equation}
where the (matrix) functions $a,b$ are linear combinations of the perturbations $u_A, v_A$. The explicit form of these relations are not particularly illuminating, so we will not report them here.

This operator can be written in terms of the (free) quasi-particle fields by using explicitly the Bogoliubov transformations\footnote{Actually, these are the inverse of the transformations considered in \ref{subsec:genquasip}.}
\begin{equation}
 \hchi_{A}(x) = \int d^3 y \left( r_{AB}(x,y) \hphi_{B}(y) +s_{AB}(x, y) \hphi^{\dagger}_{B}(y)  \right) 
\end{equation}
and the Hermitian conjugates. It will suffice to look at the first term, since the others can be treated in the same way. So:
\begin{equation*}
\hat{H}_{I,1} =
 \int d^3 x \,a^{AB}(x) \hchi_A(\xxx)\hchi_B(\xxx) +  \, h.c. = 
\end{equation*}
\begin{equation}=\int d^3y d^3 z \,\left( \mathfrak{a}^{(I)}_{AB}(y,z) \hphi_A(\yyy)
\hphi_B(\zzz)+ \mathfrak{b}^{(I)}_{AB}(y,z) \hphi_A(\yyy)^{\dagger}
\hphi_B(\zzz)+\,h.c.\,\right),
\end{equation}
where the kernels $\mathfrak{a},\mathfrak{b}$ are obtained by the convolution of the Bogoliubov kernels with the matrices $a,b$.
For example, we have:
\begin{equation}
 \mathfrak{a}^{(I)}_{AB} (\yyy,\zzz) = \int d^3 x \left[ a^{RS}(\xxx) r_{RA}(\xxx,\yyy)r_{SB}(\xxx,\zzz) +
\left(b^{RS}(\xxx) s_{RA}(\xxx,\yyy)s_{SB}(\xxx,\zzz)\right)^*\right] \,.
\end{equation}

As said, we will be interested in matrix elements between phonon states of definite momentum. Given the quadratic nature of the peculiar ``interaction term" that we are considering, the interesting matrix elements (in the Born approximation) are those between single quasi-particle states with definite momentum, \ie
\begin{equation}
 T(\hhh_C,\kkk_D) = \langle \kkk_D | \hat{H}_{int} | \hhh_C \rangle.
\end{equation}
By straightforward manipulations, we see that
\begin{equation*}
 T(\hhh_C,\kkk_D) = \int d^3y d^3z \left( \mathfrak{b}_{AB}(\yyy,\zzz) \langle 0_{qp} |\hat{b}_C(\kkk) \hphi_A(\yyy)
\hphi^{\dagger}_B(\zzz)\hat{b}^{\dagger}_{D}(\hhh)|0_{qp}\rangle \right.+
\end{equation*}
\begin{equation}+\left. \mathfrak{b}^{\dagger}_{AB}(\yyy,\zzz) \langle 0_{qp} |\hat{b}_C(\kkk) \hphi^{\dagger}_A(\yyy)
\hphi_B(\zzz)\hat{b}^{\dagger}_{D}(\hhh)|0_{qp}\rangle\right),
\label{eq:scattmat}
\end{equation}
whence
\begin{equation}
  T(\hhh_C,\kkk_D) = \int d^3 y d^3 z \left(\mathfrak{b}_{CD}(\yyy,\zzz) e^{-i\kkk\zzz} e^{i\hhh\yyy} +
\mathfrak{b}^{\dagger}_{DC}(\yyy,\zzz) e^{-i\kkk \yyy}e^{i\hhh\zzz} \right) + \mathcal{B} \delta_{DC} \delta^3(\kkk-\hhh).
\end{equation}
In this last equation we notice the appearance of a forward term with a coefficient
\begin{equation}
 \mathcal{B} = \int d^3 x \delta^{AB}\mathfrak{b}_{AB}(\xxx,\xxx),
\end{equation}
signalling the fact that a wavefunction renormalization must be performed. Notice that this coefficient is crucially dependent on the structure of the Bogoliubov coeffients but also on the perturbations $u_A,v_A$.

There are also quasi-particle pairs production/destruction graphs, corresponding to
\begin{equation}
 T(\{\hhh_C,\kkk_D\},0) = \langle \kkk_C,\hhh_D | \hat{H}_{int} | 0 \rangle,
\end{equation}
for which the other coefficients $\mathfrak{a}$ would be relevant. However, these graphs are not relevant for the discussion of Newtonian gravity, which involves only scatterings off a potential, without particle creation effects (at least not in the regimes where one usually applies the Newtonian limit, \ie well below the threshold for particle production).

Notice that, in principle,
these matrix elements can be opened up and rewritten in terms of suitable integrals of $u_A,v_B$, which are still entering linearly in these matrix elements.

The comparison with the standard Newtonian potential requires much more work with respect to the single BEC case.
We have to compare these results with the ones obtained via
\begin{equation}
 \hat{H}_{\mathrm{Newtonian}} = -\int d^3 x \sum_{A}  \mathcal{M}_A \phin(\xxx) \hphi^{\dagger}_A(\xxx)
\hphi_A(\xxx),
\end{equation}
where $\mathcal{M}_A$ are the masses of the quasi-particles.
In particular, one has that
\begin{equation}
 T_N(\hhh_C,\kkk_D) = -\delta_{CD}\mathcal{M}_C \int d^3x \phin(\xxx) e^{i(\hhh-\kkk)\xxx}
\end{equation}
\begin{equation}
 T_N(\{\hhh_C,\kkk_D\},0) =T_N(0,\{\hhh_C,\kkk_D\}) = 0 
\end{equation}

Therefore, in the multi-BEC case, the identification of the Newtonian potential involves the comparison of
all the expectation values given above with the ones given in the Newtonian theory, leading to a much less trivial
set of conditions. In particular, we want that
\begin{equation}
 T(C,D) \approx T_{N}(C,D) \propto \mathcal{M},
\end{equation}
where $\mathcal{M}$ is the mass matrix for the quasi-particles.
As one can easily see, this amounts to ask $N_c(N_c+1)/2$ conditions on the matrix elements \eqref{eq:scattmat} (for each pair of momenta!), which are ultimately related to the shape of the potential $V$ describing the condensate and to the homogeneous solution playing the role of our background. 

Since $T(C,D)$ is a Hermitian matrix, and we can extract its projection along the mass matrix $\mathcal{M}$
in such a way to identify \textit{uniquely} the candidate for the analogue of the Newtonian potential:
\begin{equation}
 - \int d^3x \phin(\xxx) e^{i(\hhh-\kkk)\xxx} = \frac{\mathrm{tr}(T_{sub}(k_A,h_B).\mathcal{M})}{\mathrm{tr}(\mathcal{M})}.
\label{eq:defNewton}
\end{equation}
where $T_{sub}$ is the transition amplitude without the forward term.

This way of manipulating the various quantities shows that the (would be) Newtonian potential is a complicated linear combination of the various potentials $\tilde{u},\tilde{v}$. Furthermore, given that the the relation \eqref{eq:defNewton} is actually momentum dependent, the relation between the Newtonian potential and the potentials $\tilde{u},\tilde{v}$ is nonlocal, when translated in real space, with the following general shape
\begin{equation}
 \phin(\xxx) = - \sum_{C,D} \frac{\mathcal{M}_{CD}}{\mathrm{tr}(\mathcal{M})} \int d^3 y  \left(\mathfrak{b}_{CD}(\xxx,\yyy) e^{-i\kkk(\yyy-\xxx)} +
\mathfrak{b}^{\dagger}_{DC}(\yyy,\xxx) e^{-i\kkk(\yyy-\xxx)} \right).
\label{eq:NewtonKernels} 
\end{equation}

This discussion shows how complicated is the identification of the analogue of a single Newtonian potential in the multi-BEC case, among the various possible candidates (after all, we should expect from the very beginning a sort
of multi-metric structure, whence a multi-potential theory).

\section{Source term}\label{sec:source}
So far the source term has been neglected. However, it is interesting to spend some time to study it, because it will lead to additional properties of the emergent theory. As we have already said, the source terms appearing in the Bogoliubov--de Gennes equations have the following structure, at the lowest order:
\begin{equation}
\Sigma_{A} = \left(\frac{\partial X_{AB}}{\partial \psi_{C}} \right)_{\psi=\bpsi} \mmm_{BC} +
\left(\frac{\partial X_{AB}}{\partial \psi^{*}_{C}} + \frac{\partial Y_{AB}}{\partial \psi_{C}} \right)_{\psi=\bpsi} \nnn_{BC} +
 \left(\frac{\partial Y_{AB}}{\partial \psi^{*}_{C}} \right)_{\psi=\bpsi} \qqq_{BC}.
\end{equation}
where all the expectation values are written in terms of the fields $\hchi$. 

As in the case of the single BEC, these terms contain vacuum contributions related to the depletion factor \cite{blackbook,FetterWalecka}, which however has a more rich structure, in this case.
Inserting the Bogoliubov transformations, we get that the expectation values in the Fock vacuum for quasi-particles are:
\begin{equation}
\mmm^{vacuum}_{AB}(x) = \int d^{3}y d^{3}z \, r_{AC}(x,y)s^*_{BD}(x,z)\langle \hphi_{C}(y) \hphi_{D}^{\dagger}(z) \rangle,
\end{equation} 
whence, using the commutation properties between the fields $\hphi$, one gets:
\begin{equation}
\mmm^{vacuum}_{AB}(x) = \int d^{3}y\sum_{C} r_{AC}(x,y)s^*_{BC}(x,y).
\end{equation}
Similarly:
\begin{equation}
\nnn^{vacuum}_{AB}(x) = \int d^{3}y\sum_{C} s^{*}_{AC}(x,y)s_{BC}(x,y), \qquad \qqq_{AB}^{vacuum}(x) \int d^{3}y\sum_{C} r^{*}_{AC}(x,y)s_{BC}(x,y).
\end{equation}
These terms are generating shifts in all the potentials due to vacuum effects (the inequivalence between the Fock vacuum for the atoms and the one for the quasi-particles). In some sense these terms migh be interesting because they could be related to small cosmological constant, as suggested in \cite{BEC}. However, here we will neglect the effect of quantum vacuum fluctuations and consider the effect of quasi-particles.

Coming back to the source term, it is obvious that the interesting parts involve states with some quasi-particles.
The expectation values to be computed are of the form:
\begin{equation}
\mmm = \langle n| \hchi_{A} \hchi_{B} | n\rangle ,
\end{equation}
where $|n\rangle$ is a stenography for an $n-$quasi-particle state. Similar expressions hold for $\nnn,\qqq$.
To be concrete, let us study, as in \cite{BEC}, the case of a state like:
\begin{equation}
 |\zeta\rangle = \sum_{A} \int d^3 x \, \zeta_A(x) \hphi_A^{\dagger}(x) |\Omega \rangle.
\label{eq:particles}
\end{equation}
This state corresponds to some quasi-particles which have a certain number density specified by the ``wavefunctions'' $\zeta_A(\xxx)$ (we are using the normalization $\langle \zeta|\zeta\rangle =1$).
Denoting with $n_A(\xxx)$ the number density of quasi-particles of color $A$, it is easy to verify that
\begin{equation}
 n_A(\xxx)=\langle \zeta |\hphi^{\dagger}_A(\xxx)\hphi_A(\xxx)|\zeta\rangle = |\zeta_{A}(\xxx)|^2 .
\end{equation}
In order to reproduce the correct Newtonian behavior, we need that the source terms become function(al)s of
these number densities. However, this is not the case. Consider, for instance, the term:
\begin{equation}
 \mmm_{AB}[\zeta] =  \langle \zeta|\hchi_A(\xxx) \hchi_B(\xxx)|\zeta \rangle .
\end{equation}
Using the definition \eqref{eq:particles}, the Bogoliubov transformations and the commutation relations between the quasi-particle fields $\hphi,\hphi^{\dagger}$, one obtains:
\begin{equation}
\mmm_{AB}[\zeta] = \mmm_{AB}^{vacuum}+ \sum_{CD} \int d^3y \,d^3 z \, \Lambda_{ABCD}(\xxx;\yyy,\zzz)\zeta^{*}_C(\yyy)
\zeta_D(\zzz),
\end{equation}
where the kernel $\Lambda$, related to the Bogoliubov coefficients, is:
\begin{equation}
 \Lambda_{ABCD}(\xxx;\yyy,\zzz) = r_{AD}(\xxx,\zzz) s_{BC}(\xxx,\yyy) + s_{AC}(\xxx,\yyy)r_{BD}(\xxx,\zzz).
\end{equation}

Similar results are obtained upon examination of the other terms obtained from $\nnn,\qqq$. The most important lesson that must be taken from these computations is that the source term is a complicated nonlocal functional of the quasi-particle distribution, not in terms of the number density of quasi-particles (as one naively would expect), but rather in terms of their quantum wavefunction \eqref{eq:particles}. 

While this is of course a crucial difficulty in obtaining some meaningful source term related directly to the number density, this feature should not come as unexpected. In fact, quasi-particles are collective motions: this represents an interesting example, albeit very simple, of mismatch between microlocality and macrolocality \cite{wroclaw}. Without any additional ingredient able to hide it, the equations of motion for the effective theory of perturbations around homogeneous background will be displaying explicitly nonlocal effects.
This feature was already present in the case of the single BEC. In the case of several components, it is easy to see that the effect is worsened by the appearance of off-diagonal terms $\zeta^{*}_{A}(\yyy)\zeta_{B}(\zzz)$, with $A\neq B$, which do not have any sort of interpretation in terms of single components.

\section{Emergent gravitational dynamics?} \label{sec:emergent}

Up to now we have already listed a number of difficulties that one has to face in order to build successfully a model of emergent Newtonian gravity starting from a multi-component BEC. Yet, we do not have addressed a most important
point: the equation of motion for the ``gravitational'' field.
 
In section \ref{sec:beyondGP}, examining the properties of the matrix $T$ encoding the scattering
amplitudes of the phonons off the inhomogeneities, we have seen where the ``gravitational'' potential is hidden, \ie in the part proportional to the mass matrix of the quasi-particles.
From our perspective, to recover something resembling Newtonian gravity, we cannot only require that all the remaining matrix elements of $T$ must be small, but also that this ``Newtonian potential'' must obey a Poisson-like
equation,
\begin{equation}
\left(- \nabla^2 + \frac{1}{L^2}\right) \phin \propto \rho(\xxx),
\end{equation}
with $L$ a constant determining the range of the interaction.
Since $\phin$ is a nonlocal linear combination of the potentials $\tilde{u},\tilde{v}$, as we have previously mentioned,
the equation of motion for $\phin$ will be obtained by considering the equations of motion for these potentials.
We have established that:
\begin{equation}
 \phin(x) = \int d^{3}y \sum_{A=1}^{N_c} \left( \alpha_A (\xxx,\yyy)\tilde{u}_A(\yyy) + \beta_A(\xxx,\yyy) \tilde{v}_A (\yyy) \right),
\end{equation}
where the precise structure of the kernels $\alpha,\beta$ can be computed from \eqref{eq:NewtonKernels}.
Taking the Laplacian of this expression\footnote{We assume also that the kernels are smooth enough to be allowed to derive under the integral.}, we obtain that:
\begin{equation}
 \nabla^2 \phin(\xxx) = \int d^{3}y \sum_{A=1}^{N_c} \left( \nabla^2_{\xxx}\alpha_A (\xxx,\yyy)\tilde{u}_A(\yyy) + \nabla^{2}_{\xxx}\beta_A(\xxx,\yyy) \tilde{v}_A (\yyy) \right), \label{nabla2}
\end{equation}
where we have explicitely introduced the notation $\nabla^2_{\xxx}$ to remember that, when acting on the kernels $\alpha, \beta$, the operator should act on the $\xxx$ variable.
This result shows that in order to get to some form of Poisson-like equation we have to unveil the structure of the kernels $\alpha,\beta$, which, in turn, are determined by equation \eqref{eq:NewtonKernels}.

By comparing \eqref{nabla2} with the desired result,
\begin{equation}
 \nabla^2\phin(\xxx) = \frac{1}{L^2} \phin + c_{A}\Sigma_{A}^{Re} + d_{A}\Sigma_{A}^{Im},
\end{equation}
we see that the kernels must not be generic. They must obey some integral identity of the form
\begin{equation}
 \int d^3 y \nabla^2_{\xxx}\alpha_{A}(\xxx,\yyy)f = \tau_A \int d^3 y \alpha_{A}(\xxx,\yyy) f + \tilde{c}_{A}
\int d^3 y \alpha_{A}(\xxx,\yyy) \nabla^2_{\yyy} f. 
\end{equation}
Albeit already very stringent, this condition is not enough to ensure that the equation will have a structure like the one we want. To see this, let us assume, for simplicity, that the kernels are proportional to Dirac distributions,
\ie that the Newtonian potential is a simple linear combination of the potentials $\tilde{u},\tilde{v}$.

\begin{equation}
 \phin(x) = \sum_{A=1}^{N_c} \left( \alpha_A \tilde{u}_A(\xxx) + \beta_A \tilde{v}_A (\xxx) \right).
\end{equation}
In this case, the action of the Laplacian is manifest:
\begin{equation}
\nabla^2 \phin(x) = \sum_{A=1}^{N_c} \left( \alpha_A \nabla^2\tilde{u}_A(\xxx) + \beta_A \nabla^2\tilde{v}_A (\xxx) \right),
\end{equation}
whence, using equations \eqref{eq:pseudopots} (and analogous for the $\tilde{v}$), we get:

\begin{equation}
\nabla^2 \phin(x) = \sum_{A=1}^{N_c} \left( \alpha_A \frac{1}{L_{A,(\tilde{u})}^2}\tilde{u}_A(\xxx) + \beta_A \frac{1}{L_{A,(\tilde{v})}^2}\tilde{v}_A (\xxx) \right)+\,\source. \label{eq:poissongen}
\end{equation}

However, unless the $L_{A,\tilde{u}},L_{A,\tilde{v}}$ are all the same, we are not able to reconstruct the needed
$\phin$ to obtain a Yukawa-mass-like term. This implies that, even if we discard the corrections due to the intrinsic nonlocality of the Newtonian potential, we have to tune the constants determining the $L_{A}$ for the $\tilde{u}_{A},\tilde{v}_{B}$, in order to make them to have the same value.

Besides the important difficulties concerning the left hand side of the Poisson equation, another important issue involves the emergence of a unique $G_{N}$, or, equivalently of some sort of universal law of gravitation. To understand whether this can happen, the equations for the potentials must be reconsidered with care, and in particular the sign of the coefficients that will lead to the source terms for the Poisson equation. 

As it is easy to understand, by doing all the diagonalizations/Bogoliubov transformations etc. we cannot control and ensure that all the coefficients in front of the pseudo-densities of quasi-particles have the same value (enforcing a universal law of gravitation). In fact, as we have already noticed, we do not have any control on the sign of these coefficients (\ie on the sign of $G_{N}^{A}$ for each component). In other words, we could have some form of antigravitation, at least in principle. Furthermore, as we have noticed in the previous section, we can have in the source term some sort of cross-terms which depend simultaneously on the distributions of quasi-particles of different colors.

\section{The $S_{\nc}$ case}\label{sec:sn}

The results of the previous section seem to be pretty discouraging, confirming the naive intuition that without
Lorentz invariance and without any form of equivalence principle imposed a priori on the microscopic theory, the effective theory of phonons will not display these properties at all.

In this section we will discuss an example in which the presence of an internal discrete symmetry ameliorates the situation, leading quite naturally to a form of emergent equivalence principle. Lorentz invariance, as usual, is not completely under control, and a tuning of the microscopic parameters has still to be performed. Nonetheless, in absence of this tuning, the excitations will be divided into multiplets which are homogeneous with respect to spacetime symmetries and to the coupling to what will turn out to be the only possible long range potential in the model.

The system we are going to discuss is the immediate generalization of the $\mathbb{Z}_2$ model that we have considered in section \ref{sec:2bec} to the case of $\nc$ components. The Hamiltonian for the system will be:
\begin{equation}
 \en \hpsi_{A} = \left( \kin - \mu \right) \hpsi_{A} + \kappa |\hpsi_A|^2 \hpsi_{A} + 
\left( \sigma \sum_{B\neq A} |\hpsi_{B}|^2 \right) \hpsi_{A} - \lambda \sum_{B\neq A} \hpsi_{B}.
\end{equation}
where $A,B=1,...,\nc$ are labels for the bosonic components and $\mu,\kappa,\sigma,\lambda$ the only four free parameters.
This Hamiltonian is invariant under the full permutation group of the components, $S_{\nc}$, and under global $U(1)$ symmetry $\hpsi_{A} \rightarrow e^{i\theta} \hpsi_{A}, \forall A$. This Hamiltonian allows the largest symmetry group leading to quasi-particles having dispersion relations with relativistic regimes. Had we imposed a $U(\nc)$ symmetry, this would not have been the case.

The Gross--Pitaevski equations read:
\begin{equation}
 \en \psi_{A} = \left( \kin - \mu \right) \psi_{A} + \kappa |\psi_A|^2 \psi_{A} + 
\left( \sigma \sum_{B\neq A} |\psi_{B}|^2 \right) \psi_{A} - \lambda \sum_{B\neq A} \psi_{B},
\end{equation}
and the $\snc$-invariant constant solution is
\begin{equation}
 \psi_{A} = n^{1/2}, \qquad \forall A,
\end{equation}
where $n$ is the number density of the condensates and the phases are set to zero, by using the global $U(1)$ symmetry. Plugging this {\it ansatz} in the GP equation we get
\begin{equation} 
 n = \frac{\mu + \nmuno \lambda }{\kappa + \nmuno \sigma}.
\label{eq:numbernc}
\end{equation}

\subsection{Quasiparticles}
The quasi-particle spectrum has to be obtained applying the method of Bogoliubov transformations to the particular case.
Splitting the operator $\hpsi_{A}$ using the mean field approximation,
\begin{equation}
 \hpsi_{A} = n^{1/2} + \hchi_{A}.
\end{equation}
The linearized field equation for $\hchi$ reads:
\begin{equation}
 \en \hchi_{A} = - \left( \kin - \mu \right) \hchi_{A} + 2 \kappa n \hchi_{A} + \kappa n \hchi_{A} ^{\dagger} + \nmuno \sigma n \hchi_{A} + \sigma n \left( \sum_{B\neq A} (\hchi_{B} + \hchi_{B}^{\dagger}) \right) - \lambda \sum_{B\neq A} \hchi_B .
\label{eq:fluctn}
\end{equation}
Going in momentum space, and defining the isovector $\isovn(k) =  \left(
 \ha_1 ,
\cdots ,
\ha_{\nc},
\had_1,
\cdots,
\had_{\nc}
 \right)\trans,$
then Equation \eqref{eq:fluctn} can be written in matrix form:
\begin{equation}
 \en \isovn(k) = \blom{A}{B}{-B}{-A} \isovn,
\end{equation}
where we have defined the matrices
\begin{equation}
 A(k) = a(k) \mathbb{I}_{\nc} + b \, \uno_{\nc}; \qquad B = c \mathbb{I}_{\nc} + d\, \uno_{\nc},
\end{equation}
with $\mathbb{I}_{\nc}$ being the $\nc\times \nc$ identity matrix, $\uno_{\nc}$ the $\nc\times \nc$ matrix having all the entries equal to one, and the four numerical coefficients given by:
\begin{equation}
 a(k) = \frac{\hbar^2 k^2}{2m} - \mu + 2 \kappa n + \nmdue \sigma n + \lambda, \qquad b= \sigma n - \lambda, \qquad c =(\kappa - \sigma) n, \qquad d= \sigma n.
\end{equation}
The matrix $\uno$ is real, symmetric and hence can be diagonalized by means of an orthogonal transformation $R$.
It is easy to see that its diagonal form is
\begin{equation}
 R^{T} \uno R = J = \left( \begin{array}{cc}
 \nc & 0 \,\,\, ... \,\,\, 0\\
0  &  \\
\vdots & \mathbf{0} \\
0 & 	
\end{array}
 \right) \label{eq:diagonaluno},
\end{equation}
where the eigenvector associated to the nontrivial eigenvalue is $(1,1,1,...,1)$, when written in term of the original basis.

As a consequence, we can easily put in diagonal form \eqref{eq:fluctn}. First of all, one has to diagonalize the single blocks of the dynamical matrix by means of the transformation $R$:
\begin{equation}
 \blom{R^T}{\mathbf{0}}{\mathbf{0}}{R^T} \blom{A}{B}{-B}{-A}  \blom{R}{\mathbf{0}}{\mathbf{0}}{R} =
\blom{D_{A}}{D_{B}}{-D_{B}}{-D_{A}}, 
\end{equation}
where
\begin{equation}
 D_{A} = a(k)\mathbb{I} + b J, \qquad D_{B} = c\mathbb{I} + d J,
\end{equation}
are diagonal matrices. Finally, by permutations of rows and columns, the dynamical matrix can be put into a block-diagonal form, where the blocks on the diagonal are two by two matrices having the same structure of the dynamical matrix of the single component BEC, namely:
\begin{equation}
\left( \begin{array}{cc}
  \alpha_i(k) & \beta_i \\
-\beta_{i} & -\alpha_i(k)
 \end{array}\right)
\label{eq:twobytwodyn}
\end{equation}
and hence leading to the same steps to complete the Bogoliubov transformations. We leave the details aside, and we focus on the emerging spectrum. The eigenvalues of \eqref{eq:twobytwodyn} are easyly obtained. The spectrum of the mode is given by:
\begin{equation}
 \omega^2_i(k) = \alpha_i^2 - \beta_{i}^2.
\end{equation}
In view of the structure of the matrix $J$ we can conclude that the spectrum of quasi-particles will consist of two families.
\begin{itemize}
 \item A gapless singlet, when $\alpha(k) = a(k) + \nc b $, $\beta=c + \nc d$, with dispersion relation given by
\begin{equation}
 \omega^{2} = \frac{\hbar^4 k^4}{4 m^2} + (w + \nc b) \frac{\hbar^2 k^2}{m}.
\end{equation}
\item A multiplet of gapped modes, with dispersion relation given by:
\begin{equation}
 \omega^{2} =\frac{\hbar^4 k^4}{4 m^2} + w\frac{\hbar^2 k^2}{m} + \omega_{0}^2,
\end{equation}
all with the same rest mass, determined by $\omega_0$ (and by $w$, controlling the speed of sound).
\end{itemize}
Here, we have introduced
\begin{equation}
 w = -\mu + 2 \kappa n + \nmdue \sigma n + \lambda
\end{equation}
to simplify slightly the expressions and highlight the relevant contributions.
Clearly, in order to have the emergence of low energy Lorentz invariance, we have to impose a constraint on the coefficients of the terms quadratic in $k$ in the two dispersion relations. In particular, the monometricity condition reads
\begin{equation}
 b = 0 \qquad \Leftrightarrow \qquad \sigma n = \lambda.
\end{equation}
This condition leads straightforwardly to a number of simplification. First of all, using \eqref{eq:numbernc}:
\begin{equation}
\sigma = \frac{\lambda \kappa}{\mu} = x \kappa, 
\end{equation}
where we have introduced the dimensionless ratio $x=\lambda/\mu=\sigma/\kappa$. Then, plugging this result back in \eqref{eq:numbernc} we have:
\begin{equation}
 n = \frac{\mu}{\kappa}.
\end{equation}
Notice that all these results match with what we have already seen in the $\mathbb{Z}_2$ case, in section \ref{sec:2bec}.

It is worth stressing again that monometricity is not imposed automatically by the underlying symmetries of the system, but it must inserted by hand with a suitable tuning of the parameters. However, the modes are organized in multiplets (with respect to the residual symmetry $S_{\nc -1}$) within which the low energy emergent metric structure is the same. Lorentz violating effects, in this kind of scenarios, will be present only when confronting different multiplets. This is perfectly analogous to the way in which Lorentz violating terms percolate in a field theory with spontaneous symmetry breaking \cite{HiggsFinsler}, and confirms the expectation that internal (custodial) symmetries might ameliorate the naturalness problems of Lorentz violating theories \cite{LiberatiNaturalness, Mattingly}. 

Before going on, we need to introduce the healing length. The healing length should be computed
by examining the dynamical properties of the condensate \cite{FetterWalecka}. Here we define it by means of the
dispersion relation of the particles, as the scale at which we have the crossover between the low energy Lorentz invariant theory to the high enery Galilei invariant theory, and it is the UV scale at which the low energy hydrodynamic description of the condensate breaks down\footnote{The contribution of the quantum potential to the hydrodynamic equations obtained from the GP equation with the Madelung representation becomes relevant when the inhomogeneities in the condensate wavefunction reach the healing scale.}. It turns out to be:
\begin{equation}
 L_{h}^2 = \frac{\hbar^2}{4 m(\mu+\nmuno \lambda)} = \left(\frac{1}{1+ \nmuno x}\right) \frac{\hbar^2}{4 m \mu}.
\end{equation}
This scale will set the characteristic scale with which we will make all the comparisons, in the next section.

\subsection{The potentials}
Let us now pass to the examination of the equation of motion for the condensates, when we turn on the backreaction effects. In the quasi-static, weak field limit,  the Bogoliubov--de Gennes equation for the condensate turns out to be:
\begin{equation}
\left( \kin -\mu \right) (u_A+iv_A) + \mu(3u_A + i v_A) + \nmuno \lambda (u_A + i v_A) + \lambda \sum_{B\neq A} (u_B - i v_{B}) = \Sigma_A,
\end{equation}
where $\Sigma_A$ are given by expectation values of the fluctuation operators, as seen in the previous section.
We can put this equation in a diagonal form, by taking real and imaginary parts. Defining 
$U=(u_1,...,u_{\nc})\trans, V=(v_1,...,v_{\nc})\trans$ and $\Sigma =(\Sigma_1,...,\Sigma_\nc)\trans$, the equation can be written in matrix form:
\begin{equation}
 \left(  \kin \mathbb{I}_{\nc} + a \mathbb{I}_{\nc} + b \uno_{\nc} \right) U = \real(\Sigma),
\label{eq:potu}
\end{equation}
\begin{equation}
 \left(  \kin \mathbb{I}_{\nc} + c \mathbb{I}_{\nc} + d \uno_{\nc} \right) V = \imag(\Sigma),
\label{eq:potv}
\end{equation}
where
\begin{equation}
 a =2 \mu + \nmdue \lambda, \qquad b= \lambda, \qquad c= \nc \lambda, \qquad d = -\lambda.
\end{equation}
The range of the effective potentials can be understood immediately by diagonalizing these matrices, so that
the equations for the effective potentials decouple and get the form:
\begin{equation}
\left( -\nabla^2  + \frac{1}{L^2} \right) \tilde{u}_i = \source
\end{equation}

Using the same technology introduced in the previous subsection, we can easily determine the way in which the condensate reacts to the presence of quasi-particles, by looking at the properties of the potentials $\tilde{u}$ and $\tilde{v}$ that diagonalize \eqref{eq:potu} and \eqref{eq:potv}.
\begin{description}
 \item{a.} There is a potential $\tilde{u}_1$ coming from the $U$ which has range given by:
\begin{equation}
 \frac{1}{L_1^2} = \frac{2m}{\hbar^2} (a + \nc b).
\end{equation}
Trivial manipulations will show that $L_1 = L_h$.

\item{b.} There are $\nmuno$ potentials $\tilde{u}$ with a different length scale.
The relation with the healing length is easy to obtain:
\begin{equation}
 \frac{L_{h}^2}{L_2^2} = \frac{1+\frac{\nmdue}{2}x}{1+\nmuno x}.
\end{equation}
In the limit\footnote{This limit has a simple interpretation. Given that the mass of the massive modes is a monotonically increasing function of $\lambda$, this limit corresponds models with a large number of heavy fields.} $\nc x \rightarrow \infty$, 
\begin{equation}
 \frac{L_{h}^2}{L_2^2} \rightarrow \frac{1}{2},
\end{equation}
which signals that these potentials will be associated to short range phenomena as well.

\item{c.} There are $\nmuno$ potentials $\tilde v$ with range given by:
\begin{equation}
 \frac{1}{L_3^2} = \frac{1}{2} \frac{\nc x}{1+\nmuno x} \frac{1}{L_{h}^2},
\end{equation}
still associated to short range deformations, as one can see by inspection of the limit $\nc x\rightarrow \infty$. 

\item{d.} Most importantly, there is a long range potential, $\tilde{v}_1$, due to the fact that
one eigenvalue of the constant matrix in \eqref{eq:potv} is given by 
\begin{equation}
 \frac{1}{L_4^2} \propto c - \nc d = \nc \lambda - \nc \lambda = 0.
\end{equation}
\end{description}

Therefore, in this class of models, there will be always a long range mode, coming from the combination of the perturbations for the velocity potentials of the condensates
\begin{equation}
 \Phi = \frac{1}{N} \sum_{B=1}^{\nc} v_{B}.
\end{equation}

\subsection{Coupling to the long range potential}
We want now to understand what is the coupling between the phonons and the potentials that we have discovered so far. Instead of using the method of scattering matrix as in the general case, here we focus on the coupling with the unique long range mode that we have, setting all the other modes to zero. Trivial manipulations show that this corresponds to the particular configuration in which:
\begin{equation}
 v_{A}(\xxx) = \Phi(\xxx), \qquad \forall A = 1,...,\nc,
\label{eq:potslongrange}
\end{equation}
with the $u_{A}$ set to zero.
We have now to reconsider the derivation of \eqref{eq:fluctn} with our hypothesis of Lorentz invariance of the low energy branch of the spectrum and of small inhomogeneities parametrized by \eqref{eq:potslongrange}.
One gets:
\begin{equation}
 i \hbar \hchi_{A} = \left( \kin - \mu \right) \hchi_A + 2 \mu \hchi_{A} + \mu \hchi_{A}^{\dagger} + 2i \mu \Phi \hchi_{A}^{\dagger} + 2i\lambda \Phi \sum_{B\neq A} \hchi_{B}^{\dagger}
\end{equation}
The coupling of the modes with the field $\Phi$ can be understood very easily, by making use of all the algebraic tricks that we have seen by now. We can rewrite the last equation in matrix form, as before. The dynamical matrix will have the form
\begin{equation}
 \blom{A}{B+iC}{-B-iC}{-A},
\end{equation}
where $A,B$ are defined as in the homogeneous case, and 
\begin{equation}
 C = 2\Phi(x)((\mu-\lambda)\mathbb{I} + \lambda \uno).
\end{equation}
Therefore, before the last Bogoliubov transformation, the dynamical matrix can be written in a block diagonal form in which each block has the following form:
\begin{equation}
\left(
 \begin{array}{cc}
\alpha_j(k) & \beta_{j} + i q_j \Phi(x)\\
-\beta_{j} - i q_j \Phi(x) & -\alpha_j
 \end{array}\right),
\end{equation}
where the value of $q_j$ depends on which eigenvector (and hence which quasi-particle) we are considering. 
It is clear, then, that the modes will be coupled to $\Phi$ in a non-universal way. The gapless singlet will have a value of the charge $q=2(\mu+\nmuno\lambda$ which will be different from the value of $q=2\mu$ for the massive quasi-particles. 

If we want to interpret $\Phi$ as a form of Newtonian gravitational potential, this is an instance in which even after the imposition of Lorentz invariance, the equivalence principle is still violated. This is of course an expected result, even considering that one of the quasi-particle species is gapless.
However, if we limit ourselves to just the gapped modes, which are in fact the ones relevant for the discussion of a Newtonian limit of gravity, within this sector all the modes are coupled in the same way to the potential, and hence they will fall in the same way under its influence.

In fact, we can push the analysis a bit forward and consider also how the quasi-particles are source for these potentials. First of all, let us consider the two classes of \vev s that we have introduced in section V (the third is not there, in this case). 
\begin{equation}
 \mmm_{AB} = \langle \hchi_{A} \hchi_{B} \rangle, \qquad \nnn_{AB} = \langle \hchi^{\dagger}_{A} \hchi_{B} \rangle.
\end{equation}
By means of the rotation matrix $R$, as we have seen, we can define new fields $\hat{\xi}$ 
\begin{equation}
 \hxi_A(\xxx) = (R\trans)^{B}_{A} \hchi_B,
\end{equation}
for which the equation of motion is block diagonal, and for which the Bogoliubov transformation can be done separately:
\begin{equation}
 \hxi_{A}(\xxx) = \int d^3k \left( \alpha_{A}(k) \hb_A(k) e^{i\kkk\cdot \xxx} + \beta_{A}(k) 
\hbd_{A}(\kkk) e^{-i\kkk\cdot \xxx} \right),
\end{equation}
where $\hb,\hbd$ are quasi-particles creation and annihilation operators. As a consequence of the separation in blocks,
we have:
\begin{equation}
 \mmm_{AB} = \sum_{C,D}\, (R\trans)^{C}_{A}\, (R\trans)^{D}_{B} \langle \hxi_{C}\hxi_{D} \rangle = \sum_{C,D} \,(R\trans)^{C}_{A} (R\trans)^{D}_{B} 
\int d^3 k \alpha_{D}(k)\beta_D(k) \langle n^{q.p.}_{D}(\kkk) \rangle \delta_{CD}
\end{equation}
where $\langle n_D^{q.p.}(\kkk)\rangle$ is the \vev~of the number of quasi-particles of (effective) color $D$ with momentum $\kkk$.
Notice that in this expression we have the same nonlocal behavior that emerged in the analysis presented in \cite{BEC}, due to the presence of the scale dependent Bogoliubov coefficients $\alpha(\kkk),\beta(\kkk)$.
We do not need to compute the form of $\nnn_{AB}$, given that we are interested in $\Phi$, which is a linear combination of $v_i$, whose equation involves the imaginary part of the source term.
It is immediate to get:
\begin{equation}
 \kin \Phi = \sum_{A} (R\trans)^{A}_{1} \Im(\Sigma_{A}) 
\end{equation}
Now, in this particular case, we know that:
\begin{equation}
 \Phi = \sum_{A} \, (R\trans)^{A}_{1} v_{A} = \frac{1}{\nc} \sum_{A} v_{A}, 
\end{equation}
whence we obtain:
\begin{equation}
 (R\trans)^{A}_{1} = \frac{1}{\nc}, \qquad \forall A,
\end{equation}
which implicitly tell us that the quasi-particles will be a source for the field $\Phi(\xxx)$ in a universal way. When
we introduce all the required dimensionfull quantities to establish the relation between the gravitational field and the field $\Phi(\xxx)$, we will obtain that the analogue of the Newton's constant $G_{N}$ is particle independent, at least in the massive sector.

\section{Conclusions and outlook}
The analysis presented in this paper is a (partial) assessment of the idea that gravity is a hydrodynamic description of some pre-geometric degrees of freedom, at least for what can be inferred by considering
a BEC analogue model\footnote{A vaguely similar situation was considered in \cite{consoli}, albeit the setting is pretty different, both conceptually and technically.}.
This rather long and detailed discussion shows that getting a gravitational theory as simple as Newtonian gravity within a BEC analogue model, without a strong guiding principle like gauge symmetry or a suitable custodial symmetry, is a very long shot. We have shown that, exactly as in other instances of emergent gravity \cite{nordstrom}, an underlying symmetry principle is needed in order to tame a number of features that would be in plain contrast with
current observations.

Summarizing, the emergence of a Newtonian gravitational dynamics (and low energy Lorentz invariance) out of a multi-component BEC-like system requires a number of conditions and display a number of nonconventional features:
\begin{itemize}
\item A number of ``monometricity'' conditions \eqref{eq:lineardisp}, \eqref{eq:monometricity} are needed to translate the gap in terms of mass, and also to guarantee a form of low energy Lorentz invariance.
\item Some diagonalizability conditions \eqref{eq:diagonalizability} are required to avoid mixing between potentials.
\item The ``non gravitational interactions'', \ie the part of the matrix $T(A,B)$ \eqref{eq:scattmat} which is not proportional to the mass matrix of quasi-particles, must be small (in other words we need to screen the color of the components).
\item To obtain a sensible Poisson equation we need a precise tuning of the parameters to control the ranges of the various potentials involved (see discussion after Eq.\eqref{eq:poissongen}).
\item In order to get a single Newton's constant we need additional tuning.
\item Nonlocality is heavily affecting the gravitational sector: given the mismatch between particle's and quasi-particle's notion of locality, the effective equations of motion are displaying nonlocalities at all the possible levels.
\end{itemize}
These are, in extreme summary, the results of sections \ref{sec:general}--\ref{sec:emergent}. Besides them, we have also shown that under certain conditions we can have some interesting situations in which things seem to work a bit better, by considering a model in which an additional internal (discrete) symmetry is imposed on the system. This
leads to a series of nice features:
\begin{itemize}
 \item (partial) taming of Lorentz violation: particles are organized in multiplets, when their dispersion relations are considered;
\item a long range ``gravitational potential'' appears;
\item the quasi-particles are coupled to this gravitational potential in a pattern according to which the equivalence principle (in the sense of universality of free fall) is implemented within multiplets.
\end{itemize}

With respect to these results, the case of the single component BEC seems peculiar, an oversimplified case just as the single field normal mode analysis \cite{singlemode} is too simple with respect to the multi-field case to get an emergent spacetime \cite{normalmodes}: many of the phenomena proper of multicomponent system cannot be displayed because there are too few components. However, the multi-component case has also the potentiality of generating effective long range interactions, and hence still the analysis is worth the pain.

The case of multicomponent systems shows that in order to produce a phenomenologically viable emergent gravity scenario, insisting in abandoning the basic notions of geometry Lorentzian geometry, together with the principles that govern the coupling to matter, one has to assume a lot of additional conditions.
Stated in a more explicit way, without having some alternative form of equivalence principle from the very beginning, we should not expect it to emerge at the end. In fact, as it has been shown in \cite{nordstrom}, it might be that there is an accidental emergent symmetry that takes care of the equivalence principle. However, in absence of a detailed full-fledged mechanism of emergent gravity from a pre-geometric theory it is impossible to address in a complete way the issue.\footnote{For a discussion on the limits on emergent gravity models (and additional references) we refer to \cite{Jenkins} and to \cite{SIGRAV}, where the role of symmetries is discussed with care.}
Of course, this difficulty with the equivalence principle is nothing else than the dynamical counterpart of the necessity of a custodial symmetry taking care of Lorentz violating terms at low energy in Lorentz violating scenarios (for a recent discussion of this point and references, see \cite{MaccioneLiberati}, sec. III C).

From the discussion of the $\snc$ model in section \ref{sec:sn}, it appears that even the imposition of an additional
internal symmetry is not able to keep under control completely the system: to enforce Lorentz symmetry at low energy we need to perform additional tunings. Furthermore, given that the gravitational field, in the model, is completely disconnected from any form of gauge principle based on the Lorentz group, even imposing Lorentz invariance is not enough to enforce an equivalence principle. From the point of view of real world, this means that if the standard model and general relativity are the low energy limits of a Lorentz violating theory, the underlying theory is not generic at all, but must have very specific symmetry properties\footnote{In this respect, it is instructive to consider the case of the recently proposed anistropic scaling model by Ho\v{r}ava \cite{Horava}. See, for instance,
\cite{ThomasHorava} for a discussion of related issues and references therein. See also \cite{HoravaSym} for a proposal of a custodial symmetry.}.

It must be said that this point of view is only partial, since it seems to be associated to the quantum/emergent gravity scenarios most commonly encountered in the analogue models literature, which are based, in one form or the other, on some condensation mechanism in a pre-existing space. It might that different classes of phenomena are needed, \ie that one should look for a completely different mechanism for the emergence of spacetime and matter fields, based on totally different concepts. In this direction, it is important to mention the cases of matrix models for 2D quantum gravity \cite{mm} and for higher dimensional spacetimes \cite{Steinacker}, as well as the Group Field Theories (GFT) \cite{GFTrev} recently proposed to implement in a background independent way the functional integral approach to gravity.

In particular, as it is suggested by matrix models, matter degrees of freedom and geometry could be encoded
in the same objects, which are pre-geometric and are not obviously associated to some fields living on some manifold. The geometric and matter degrees of freedom will appear as we know them only when the statistical system defined by matrices is forced to approach criticality \cite{mm}. A similar situation occurrs in the case of quantum graphity models \cite{graphity}.
The way in which matter fields and geometry (and their reciprocal relationship) are emerging from a pre-geometric theory has been investigated also within GFT \cite{GFTmatter}. In this case, however, the correspondence with the continuum semiclassical limit represented by general relativity is much less understood, even though some progresses concerning the appearance of a hydrodynamic limit have been done \cite{paperGFT}. In all these scenarios, the very nature of gravity and matter fields as excitations of the same underlying structure might automatically guarantee the emergence of low energy local Lorentz invariance and of a gravitational theory satisfying some form of equivalence principle. Of course, more work is needed to understand whether this is the case.

Lacking a solid derivation of general relativity from consistent pre-geometric models, we can still gather insights from analogue models. In particular, the work presented in this paper is not only elucidating the conditions under which we can have, from an emergent systems, phenomenology which is compatible with observations, but also what are the potential candidates to detect that spacetime geometry is an emergent phenomenon, in this case Lorentz violation, nonlocal phenomena in the various sectors of the theory and possible violation of equivalence principle\footnote{Of course, we have to agree, within a relativistic theory, which form of the principle we want to assess.}.

In this paper we have considered the very simple case of nonrelativistic scalar bosons. 
However, it is natural to ask what happens in a system with
several fermionic fields, and under which condition we can have the emergence of gauge interactions, besides the gravitational potential. A model like this would be much closer to the kind of physics we are familiar with from high energy physics, and allow us to discuss even additional features like the appearance of chirality. Of course, this requires us to change completely the class of models (and look for Fermi liquids as in \cite{volovik}). We leave this for future investigations.

\acknowledgments
I would like to thank F.~Girelli and S.~Liberati for discussions and comments on earlier versions of this paper.

\end{document}